\documentclass[preprint2]{aastex}

\def\ltsima{$\; \buildrel < \over \sim \;$}
\def\simlt{\lower.5ex\hbox{\ltsima}}
\def\gtsima{$\; \buildrel > \over \sim \;$}
\def\simgt{\lower.5ex\hbox{\gtsima}}
\def\cgs{{erg cm$^{-2}$ s$^{-1}$}}

\def\ergs{{erg s$^{-1}$}}
\def\cm2{{cm$^{-2}$}}

\def\xrd{{$\chi^{2}_{\rm \nu}$/dof}}
\def\xddof{{$\chi^{2}/$dof}}

\def\fhx{{$F_{\rm 2-10}$}}

\def\lums{{$L_{0.5-2 keV}$}}
\def\lum{{$L_{2-10 keV}$}}

\def\lx{{$L_{X}$}}

\def\xmm{{\em XMM--Newton}}
\def\chandra~{{\em Chandra}}
\def\sax{{\em BeppoSAX}}

\def\nhgal{{N$_{\rm H}^{\rm Gal}$}}

\def\chandra{{\em Chandra}}

\def\xmm{{\em XMM--Newton}}
\def\nhgal{{N$_{\rm H}^{\rm Gal}$}}
\def\nh{{N$_{\rm H}$}}
\def\epic{{\em EPIC}}

\def\f14{{10$^{-14}$}}
\def\f13{{10$^{-13}$}}
\def\f12{{10$^{-12}$}}
\def\f11{{10$^{-11}$}}
\def\e22{{10$^{22}$}}

\def\feka{{Fe K$\alpha$}}
\def\ir{{IRAS 20210$+$1121}}
\def\irs{{I20210S}}
\def\irn{{I20210N}}

\def\lo{{$L_{\rm [OIII]}$}}

\shorttitle{XMM-Newton discovery of an AGN pair in the interacting galaxy IRAS 20210$+$1121}
\shortauthors{Piconcelli et al.}

\begin{document}

\title{Witnessing the key early phase of quasar evolution:\\ an obscured AGN pair in the interacting galaxy IRAS 20210+1121}

\author{Enrico Piconcelli\altaffilmark{1}, Cristian Vignali\altaffilmark{2}, Stefano Bianchi\altaffilmark{3}, Smita Mathur\altaffilmark{4}, Fabrizio Fiore\altaffilmark{1}, Matteo Guainazzi\altaffilmark{5}, Giorgio Lanzuisi\altaffilmark{6}, Roberto Maiolino\altaffilmark{1}, Fabrizio Nicastro\altaffilmark{1,7,8}}

\altaffiltext{1}{Osservatorio Astronomico di Roma (INAF), Via Frascati 33, I-00040 Monte Porzio Catone (Roma),  Italy; enrico.piconcelli@oa-roma.inaf.it}
\altaffiltext{2}{Dipartimento di Astronomia, Universit\`a  di Bologna, Via Ranzani 1, I-40127 Bologna, Italy}
\altaffiltext{3}{Dipartimento di Fisica, Universit\`a  degli Studi Roma Tre, via della Vasca Navale 84, I-00146 Roma, Italy}
\altaffiltext{4}{Ohio State University, 140 West 18th Avenue, Columbus, OH 43210, USA}
\altaffiltext{5}{European Space Astronomy Centre of the European Space Agency, PO Box 78, Villanueva de la Ca\~{n}ada, E-28691 Madrid, Spain}
\altaffiltext{6}{IASF (INAF), via del Fosso del Cavaliere 100, I-00133 Roma, Italy}
\altaffiltext{7}{Harvard-Smithsonian Center for Astrophysics, 60 Garden Street,  MS-04, Cambridge, MA 02155, USA}
\altaffiltext{8}{IESL, Foundation for Research and Technology, 711 10, Heraklion,  Crete, Greece}

\begin{abstract}
We report the discovery of an active galactic nucleus (AGN) pair in the interacting galaxy system IRAS 20210$+$1121 at $z$ = 0.056. An {\it XMM-Newton} observation reveals the presence of an obscured (\nh\ $\sim$ 5 $\times$ 10$^{23}$ \cm2), Seyfert-like ($L_{2-10 keV}$ = 4.7 $\times$ 10$^{42}$ \ergs)
nucleus in the northern galaxy, which lacks unambiguous optical AGN signatures.
Our spectral analysis also provides strong evidence that the IR-luminous southern galaxy hosts a Type 2 quasar  embedded in a bright starburst emission. 
In particular,  the X-ray primary continuum from the nucleus appears totally depressed in the {\it XMM-Newton} band as expected in case of  a Compton-Thick  absorber, and only the emission produced by Compton scattering (``reflection'') of the continuum from circumnuclear matter is seen.
As such,  IRAS 20210$+$1121 seems to provide an excellent opportunity to witness a key, early phase in the  quasar evolution predicted by the theoretical models of quasar activation by galaxy collisions.

\end{abstract}


\section{Introduction} 
Current supermassive black hole
(SMBH) formation and SMBH/galaxy co-evolution models predict an early
dust-enshrouded phase  associated with rapid SMBH growth 
triggered by multiple galaxy encounters
\citep{silk98,dimatteo05,hopkins08}.   Tidal interactions 
favor both violent star formation as well as  funneling of
large amount of gas into the nuclear region to feed (and obscure)  the
accreting SMBH \citep[e.g.,][]{urrutia08, sha10}.  The importance of mergers
increases with redshift \citep{conselice03, lin08} and 
their fundamental role at the peak epoch of luminous AGN (i.e. quasar)
and intensive star-formation activity 
at 1.5\simlt\ $z$ \simlt\ 3 is widely accepted.
 
Over the last few years, an increasing number of interacting and
disturbed molecular gas-rich galaxy systems showing both coeval
powerful starburst (SB) and quasar activity at high $z$ have indeed
been unveiled \citep[e.g.,][]{carilli02,dasyra08}. The counterparts in the local
Universe to such luminous, high-$z$ mergers are the (ultra)-luminous
infrared ($L_{\rm IR}$ $>$ 10$^{11}$ $L_{\odot}$) galaxies,
i.e. (U)LIRGs \citep{sanders96}. 
 In
particular, these powerful objects should provide the opportunity for
probing an inevitable outcome of the hierarchical merging process,
i.e. the existence of dust-enshrouded double/multiple SMBHs within the
envelope of the host galaxy merger \citep{colpi08}.   Despite being
widely pursued,  direct observational evidence for AGN pairs in ULIRGs
(as well as in all of the other types of galaxies) has been very
limited so far. In the last few years X-ray observations
with arcsec angular resolution have provided one of the most efficient
tools to disclose such systems.  \citet{komossa03} discovered the
first and unambiguous example of an active SMBH pair separated by
$d$ $\sim$ 1.4 kpc in the center of the ULIRG NGC 6240.  Additional examples of dual AGNs with a close
separation were unveiled by \citet{bianchi08} and
\citet{ballo04} in the ULIRGs Mrk 463  ($d$ $\sim$ 3.8 kpc) and Arp
299 ($d$ $\sim$ 4.6 kpc), respectively,   on the basis of \chandra\
data. \citet{guainazzi05a} have reported the discovery
of an X-ray bright AGN pair in ESO590-IG066,  an early-phase ($d$ $\sim$ 10.5 kpc) merging system at $z$ = 0.03. 
 It is worth noting that
in all these cases hard ($>$2 keV) X-ray data have been
crucial to detect   activity from both SMBHs of the galaxy pair. Indeed, at least one pair member  lacks optical/IR signatures of AGN activity,  that suggests we are observing a non-standard AGN phase. 

Furthermore, a handful of kpc-scale dual AGN candidates have been
recently uncovered in galaxy mergers by the detection of
spatially-resolved,  double-peaked emission line profiles with
velocity offsets of a few hundreds km s$^{-1}$ \citep{civano10,liu10},
thus doubling the total number of {\it bona fide} active SMBH pairs
collected so far and, in turn, opening interesting perspectives for
further advances in this field of research.

Here,  we present  the discovery of an AGN pair consisting of a Compton-thick (CT) Type 2 quasar and a heavily obscured Seyfert 2-like source in the interacting galaxy IRAS 20210$+$1121 (e.g., Sect. 2). This discovery is based on the imaging and spectral analysis of \xmm\ data described in Sect. 3. We discuss our results and conclude in Sect. 4.  A cosmology with $H_0$ = 70 km s$^{-1}$ Mpc$^{-1}$,
$\Omega_\Lambda$ = 0.73 and  $\Omega_M$= 0.27 is assumed throughout \citep{spergel07}.

\section{IRAS 20210+1121}
IRAS 20210$+$1121 is  an interacting system of two galaxies separated by  12.2 arcsec \citep{arribas04, davies02}.
The larger component of the
system, i.e. the southern galaxy (\irs\ hereafter),  shows a noticeable
spiral arm structure, while the northern object (I20210N hereafter) is
more spheroidal in shape. Furthermore, a  bridge of emission
connecting both galaxies is also visible in the optical band. However,
narrow-band H$\alpha$ imaging shows  that \irs\ has a centrally concentrated, 
featureless morphology, while \irn\ is barely visible \citep{heisler95}.
 \irs\ is a LIRG
 ($L_{IR}$ = 7.8 $\times$ 10$^{11}$ $L_\odot$) with a Seyfert 2 nucleus at
$z$ = 0.056, and a [OIII] luminosity \lo\ = 2.04 $\times$ 10$^{43}$
\ergs\ \citep[e.g.,][]{perez90,shu07}. 

\irn\ has  a very faint emission line
spectrum, but sufficient to derive the same $z$  of the companion.  An important feature of the optical spectrum of \irn\ is
an intensity ratio   [NII]$\lambda$6584/H$\alpha$
$\approx$3. Such a value is typical of both LINERs and Seyfert galaxies, and  no firm
conclusion on the presence of an AGN in this source can be
drawn on the basis of these low signal-to-noise ratio  data.
 Furthermore,  a near-IR spectroscopic study of \irn\  found a featureless near-IR continuum \citep{burston01}.
\begin{figure}
\begin{center}
\includegraphics[width=8.cm,height=7.0cm,angle=0]{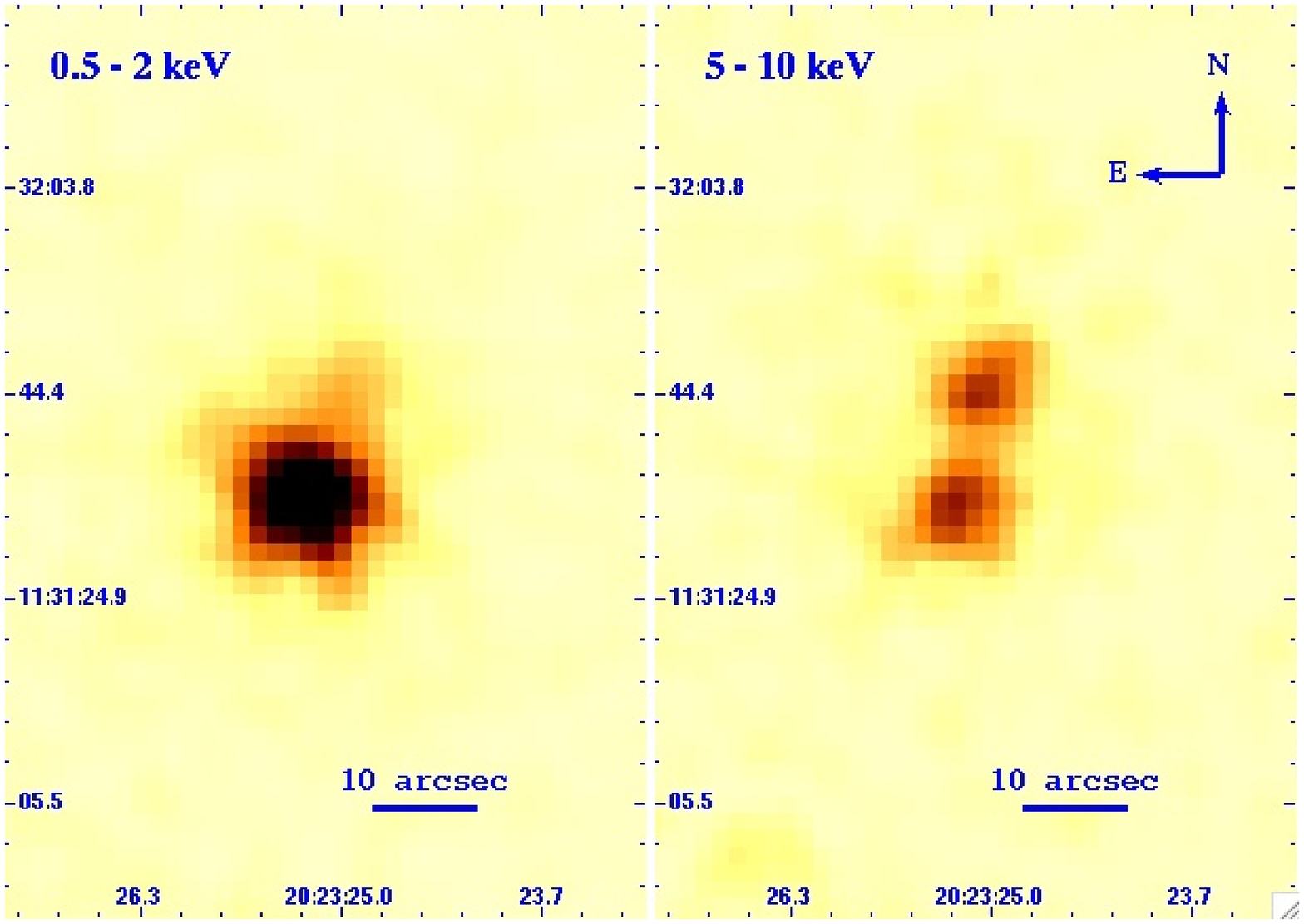}
\caption{\xmm\ \epic\ PN Gaussian-smoothed ($\sigma$ = 2 arcsec) image of the interacting galaxy system IRAS 20210$+$1121 in the 0.5-2 keV ({\it left panel}) and 5-10 keV  ({\it right panel}) energy range. }
\label{fig:12}
\end{center}
\end{figure}

\irs\  was detected by \sax\ at a 2--10 keV flux level of \fhx\ $\approx$ 2.9 $\times$
10$^{-13}$ \cgs~and  only $\sim$160 counts were collected in the
0.1--10 keV band (given the arcmin angular resolution of \sax, any
possible emission from \irn\ cannot be discerned). The analysis of
these data by \citet{ueno00} revealed  a very
flat continuum slope $\Gamma$ = 0.5$^{+0.7}_{-1.0}$ and the  remarkable
presence of a strong  (EW$_{\rm Fe}$ = 1.6$^{+2.3}_{-1.1}$ keV) 
\feka~line that  led these authors to suggest that \irs\
may host a CT AGN.  The observed \fhx\ to [O III] flux ratio of $<$ 0.1  inferred for \irs\ also hints for a CT absorber scenario \citep{guainazzi05b}.
However, it is worth bearing in mind the large
errors affecting the spectral parameters derived from this
observation. 

\section{Observations and Data Reduction}

We observed IRAS 20210$+$1121 with \xmm\  on May 22,
2009 for about 75 ks (Obs. ID.: 0600690101).  The observation was
performed with the {\it EPIC } PN and MOS cameras operating in
Full-Window mode and with the MEDIUM filter applied.  Data were reduced
with SAS v9.0 using standard procedures
and the most updated calibration files.
 The event
lists were filtered to ignore periods of high background flaring
according to the method presented in \citet{pico04} based on
the cumulative distribution function of background lightcurve
count-rates and maximization of the signal-to-noise ratio. 
 The PN source counts were extracted
from a circular region of 10 (\irn) and 8 (\irs) arcsec centered at ($\alpha_{2000}$ = 20$^{h}$23$^{m}$25.04$^{s}$; $\delta_{2000}$ = $+$11$^{\circ}$31$^{\prime}$47.7$^{\prime\prime}$) and ($\alpha_{2000}$ = 20$^{h}$23$^{m}$25.35$^{s}$; $\delta_{2000}$ = $+$11$^{\circ}$31$^{\prime}$30$^{\prime\prime}$), in order to avoid any cross-contamination
between the two regions and include the maximum number of counts with $E >$ 5 keV. For the two  MOS cameras the extraction radius was of 9 arcsec for both sources.
The background spectra were extracted from source-free, much larger circular regions on the same chip and close to the target.
After this screening, the final net exposure times in case of \irs(\irn) were 61.2(60.1) and
69.7(70) ks for PN and MOS, respectively.
 Appropriate response and ancillary files for all the \epic~cameras were
created using RMFGEN and ARFGEN tasks in the SAS, respectively.
 Spectra were rebinned so that
each energy bin contains at least 20 counts to allow us to use the
$\chi^2$ minimization technique in spectral fitting.

In this Letter we present and discuss the PN spectral results only, since this detector has a better sensitivity
over the broad 0.3-10 keV range compared to both MOS cameras (even if co-added together), and above 5 keV in particular. Nonetheless, we checked that consistent results were obtained including the MOS data in our analysis.

\begin{figure}
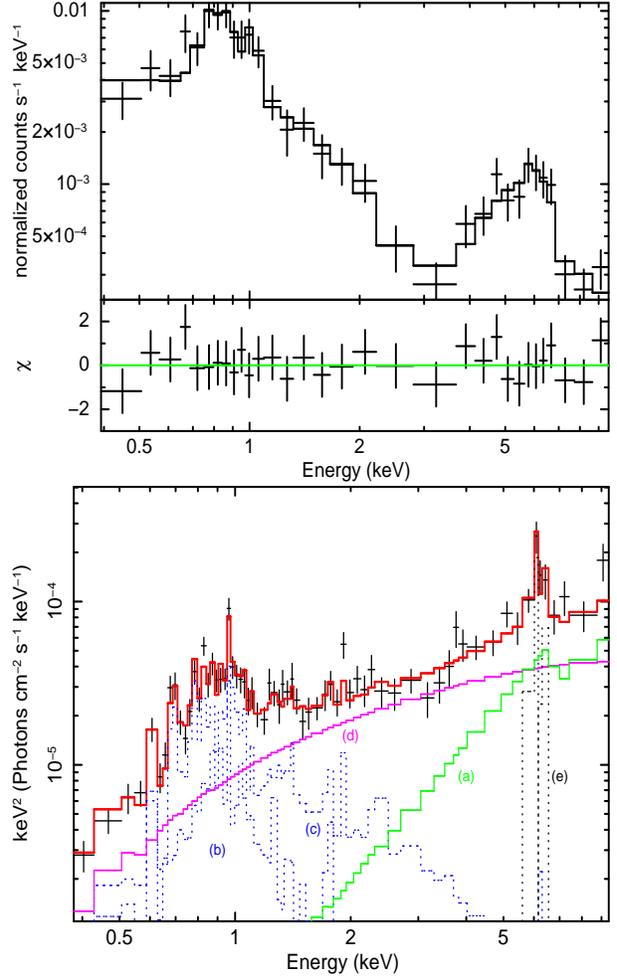

\begin{center}
\includegraphics[width=6.5cm,height=8.cm,angle=-90]{fig2.ps}
\vspace{0.1cm}\includegraphics[width=6.5cm,height=8.cm,angle=-90]{fig3.ps}
\caption{{\it Top:} 
Best-fit data and folded model (thick line), plus residuals, of the PN spectrum of \irn. {\it Bottom:}
The  PN spectrum of \irs\ with the ``composite'' CT AGN$+$SB best-fit  model  plotted as a thick line. The individual model components are also shown: (a) pure reflection continuum resulting from a power law  illumination of cold material; (b)$+$(c) thermal plasma components; (d) cutoff power-law associated with the X-ray binaries; and (e) two narrow Gaussian lines at 6.4 and 6.7 keV, respectively. See text for details.}
\label{fig:12}
\end{center}
\end{figure}

\section{Results} 

An  important result from the \xmm\ observation is presented in Fig. 1, showing
the X-ray images of \ir\ in the 0.5-2 and 5-10 keV bands.
In the soft band, only the X-ray emission centered on the [OIII]-luminous galaxy of the pair, i.e. \irs,  is clearly visible.
Whereas the very hard X-ray image reveals the presence of two sources being almost comparable in intensity, and
spatially coincident with the radio nucleus of the southern galaxy, or with the optical centroid of the northern galaxy  (once an absolute astrometry uncertainty of 2 arcsec and the dispersion of photons due to the PSF are considered\footnote{See http://xmm2.esac.esa.int/docs/documents/CAL-TN-0018.pdf for more detail.}), respectively.
At their redshift, the projected separation between the intensity peaks of both sources in the 5-10 keV image shown in Fig. 1 is  $d$ $\approx$ 11 kpc.
The simplest interpretation of the hard X-ray image is that \irn\ may also host an obscured AGN, thus revealing, in turn, the presence of an AGN pair in this interacting system. 
Direct evidence to support this buried AGN pair hypothesis comes from the X-ray spectroscopy of \irn.

The spectral
analysis of \epic~data of both sources  was carried out using
the XSPEC v12 software package.  The Galactic column
density of \nhgal\ = 9.73 $\times$ 10$^{20}$ \cm2~derived from
\citet{dick90} was adopted in all the fits. Henceforth,
errors correspond to the 90\% confidence level for one interesting
parameter, i.e. $\Delta\chi^2$ = 2.71. 

We yielded a very good description of the spectrum of \irn\ with a
typical Compton-thin Seyfert 2 model \citep[e.g.,][]{turner97}, as shown in Fig. 2 (top panel).  The primary
X-ray continuum power law is absorbed by a column density of \nh\ =
(4.7$^{+1.7}_{-1.0}$) $\times$ 10$^{23}$ \cm2\ and  exhibits a
slope of $\Gamma$ = 2.0$\pm$0.2.  The emission in the soft portion of the spectrum
(the so-called {\it soft excess} component) is well fitted by an
additional unabsorbed power law fixing its photon index to that of the
absorbed power law, but with a different normalization ($\sim$ 3\% of
the primary continuum), plus  three  narrow Gaussian emission lines.
The best-fit values for the energy of these lines are $\sim$0.82,
$\sim$0.92, and $\sim$1.07 keV, which can be identified with Fe
XVII 3d-2p, Ne IX K$\alpha$ and Ne X K$\alpha$/Fe XXI 3d-2p
transitions, respectively.
 Such a   {\it soft excess} component has
been detected in most of the obscured AGNs, and it is typically
explained as  emission
from large-scale ($\sim$0.1--1 kpc; see \citet{bianchi06}) photoionized gas, dominated by a wealth of strong
emission lines from hydrogen- and helium-like ions of the most
abundant metals, from carbon to sulfur \citep{guainazzi07,kinka02}.
Assuming this spectral model (\xddof\ = 15/20), we measured a 0.5-2 keV
flux of 1.6 $\times$ 10$^{-14}$ \cgs, and a 2-10 keV flux of 1.2 $\times$ 10$^{-13}$ \cgs.
 After correcting for absorption, this flux corresponds to a luminosity of \lum\ = 4.7 $\times$ 10$^{42}$ \ergs\ in the hard band.
Such a value of the 2-10 keV luminosity falls well within the AGN luminosity
range \citep[e.g.,][]{maiolino03}, thus providing  unambiguous evidence
for the existence of an active SMBH at the center of \irn.\\


The \xmm\ spectrum of \irs\ is very complex as shown in Fig. 2 (bottom panel). This was  expected on the basis of millimeter/IR/optical data that have revealed the simultaneous presence of star-forming and nuclear activity in this galaxy \citep{horellou95,burston01,perez90}.
In particular, from the  1.4 GHz radio(far-IR) luminosity of \irs, a star-formation rate SFR $\sim$ 120(75) M$\odot$ yr$^{-1}$ can be estimated according to the relationship reported in  \citet{ranalli03}.  
We have indeed found an excellent description of the \xmm\ data assuming a composite SB $+$ AGN emission model (\xrd\ = 0.96(65)). 
The soft X-ray SB emission has been fitted by the superposition of two thermal emission components (MEKAL model in  XSPEC) with solar metallicity and a temperature kT = 0.58$\pm$0.08 and kT = 1.25$^{+0.31}_{-0.16}$ keV, respectively, in agreement with typical values of temperature  measured in other well-known star-forming galaxies \citep{ptak99}.  The hard X-ray emission has been described assuming an absorbed cutoff-power law model in the form $E^{-\Gamma}$ exp$^{-h\nu/kT}$ with photon index $\Gamma$ = 1.1$\pm$0.2 and cutoff energy fixed to 10 keV. Such a spectral shape is expected in case of a contribution from flat-spectrum bright Low-Mass X-ray binaries and, mostly,  from High-Mass X-ray binaries (HMXBs), as pointed out by \citet{persic02}. Another striking characteristic of the  emission from HMXBs is a strong Fe XXV  emission line at 6.7 keV, that is indeed observed in the spectrum (see Fig. 2, bottom panel) with a poorly-constrained value of equivalent width EW $\sim$ 500 eV.
We derived a soft(hard) X-ray luminosity of \lums\ = 5.2(6.6) $\times$ 10$^{41}$ \ergs\ for the SB emission.
According to  \citet{ranalli03}, these values  imply  a star-formation rate SFR $\sim$ 110$-$130 M$_\odot$ yr$^{-1}$, which is consistent with the SFR values derived both from the radio and far-IR luminosity reported above.
The goodness of this match therefore lends further support to the idea that the hard X-ray power-law component originates from the population of HMXBs expected to be present in the SB regions of \irs.

Our best-fit model to the \xmm\ data includes a component due to X-ray reflection  from cold circumnuclear matter  with  \nh\simgt\ 1.6 $\times$ 10$^{24}$ \cm2~(i.e. CT), and  an \feka\ emission line to account for the reprocessed AGN emission visible in the 0.3-10 keV band, the X-ray primary continuum emission being completely blocked in this energy range \citep{ghisellini94}. 
 
The energy centroid of the line is at 6.35$^{+0.06}_{-0.04}$ keV and the EW measured with respect to the reflection continuum is  900$\pm$400 eV. These values unambiguously indicate  an origin from reflection in cold circumnuclear CT material. 
The X-ray primary continuum from the AGN is totally depressed in the {\it EPIC} band, as expected in case of  a CT absorber.
We derive an AGN flux (\fhx\ = 7.7 $\times$ 10$^{-14}$ \cgs), accounting for  
$\sim$47\% of the total (AGN$+$SB) 2-10 keV flux.
The observed \lum\ of the AGN component is 5.3 $\times$ 10$^{41}$ \ergs: according to the reflection-dominated/CT scenario it should be at most 1-2\%  of the de-absorbed luminosity \citep[e.g.,][]{comastri04,levenson06}, suggesting an intrinsic  \lum\ $>$  0.5-1 $\times$ 10$^{44}$ \ergs, which is consistent with the expectation based on the \lo.
This result matches well with the hypothesis of a CT absorbing screen and, hence,  the presence of a quasar 2 (with \lx\ $>$10$^{44}$ \ergs) at the heart of \irs. 

We also tried a model assuming a transmission scenario for the AGN emission below 10 keV. We fixed the photon index of the continuum power law to the canonical value of $\Gamma$ = 1.8 due to the limited statistics. This fit is statistically as good as the reflection-dominated fit discussed above, with a resulting \nh\ = (3.2$^{+2.0}_{-1.5}$)  $\times$ 10$^{23}$ \cm2. 
However, this \nh\  implies an \lum\ = 1.2 $\times$ 10$^{42}$ \ergs, which is two orders of magnitude lower than expected on the basis of the \lo, and an EW of the \feka\ line against the absorbed continuum of $\sim$ 250 eV \citep{ghisellini94,guainazzi05b}, while we measured an EW = 620$\pm260$ eV. These two considerations tend to disfavor the transmission scenario and lead us to assume the presence of a CT screen along our line of sight to the nucleus of \irs\ as the most likely interpretation of the \xmm\ data.

\section{Conclusions}

The \xmm\ observation of IRAS 20210$+$1121 presented here has unveiled the existence of an obscured AGN pair placed at a projected distance of $d$ $\sim$ 11 kpc in this interacting galaxy system.
In particular,
we have discovered a Seyfert 2-like AGN in the nucleus of \irn, for which neither optical nor near-IR spectroscopic observations have provided unambiguous evidence for the existence of an active SMBH at its center.  
Furthermore, the results of our spectral analysis have provided evidence that the southern member of the pair, the LIRG \irs, optically classified as a Seyfert 2 galaxy, likely hosts a powerful AGN hidden behind a CT absorber. The AGN is embedded in a strong SB emission accounting for $\sim$50\% of the 2.0-10 keV flux measured for \irs.

IRAS20210$+$1121 is therefore a rarely-observed example of CT quasar 2 plus optically 'elusive' AGN pair observed during the initial stage of the interaction between their host galaxies, that are  still easily identifiable, but also show a well-developed tidal bridge \citep{arribas04}.  
As such, this system seems to provide an excellent opportunity to witness a  merger-driven  phase of quasar fueling predicted by most of the evolutionary models based on the co-evolution of SMBHs and their host galaxies.

The mismatch between the optical/near-IR and the hard X-ray appearances of the nuclear spectrum of \irn\ can be explained in terms of a completely blocked line of sight to the nuclear region, so that the narrow line region (NLR) is also obscured \citep{maiolino03}.
The geometrical properties of the absorber should be thereby different from those assumed in typical Seyfert 2 galaxies, i.e. a pc-scale torus-like shape. The absorber in \irn\  may be characterized by a much more extended distribution (over a few hundreds of pc)
 in a way that obscures the NLR. Alternatively, it could be spherically symmetric blocking the flux of ionizing UV photons responsible for the line emission in the NLR. The AGN in \irn\ is revealed through hard X-ray observations only; this
feature is shared by most AGN pairs discovered lately thanks to hard X-ray observations \citep[e.g.,][]{komossa03,guainazzi05a,bianchi08}. An intriguing explanation for this behavior can be given in terms of a dust-enshrouded circumnuclear environment due to merger-induced processes favoring gas concentration in the galaxy center. For instance, there may be an extra-torus optical/X-ray absorber lying far from the nucleus, and  outside the NLR, being likely associated with prominent dust lanes in the disturbed host galaxy.

The discovery of a CT quasar 2 in \irs\ is in itself very important given the paucity of low$-z$ members of this peculiar class of AGN detected so far, which are considered a key ingredient in the synthesis models of the  Cosmic X-ray Background. This can be ascribed to  their low surface density and their absorption-induced faintness at the wavelengths where classical large-area surveys have been performed (i.e. optical, near-IR, UV, X-rays), that make the luminous CT AGN population extremely difficult to observe. Selection criteria based on mid-IR vs. optical colors \citep[e.g.,][and references therein]{lanzuisi09} have been proven to be efficient in discovering a large number of heavily obscured quasar candidates at $z$ $\geq$ 1. Unfortunately, most of these sources are detected at very faint flux levels (\fhx~$\ll$ 10$^{-14}$ \cgs), making an appropriate X-ray spectral follow-up extremely time-consuming.

Further support to the CT  quasar 2 nature for the AGN in \irs\ is provided by the comparison of the expected value
of  the X-ray bolometric correction $r_{X,bol}$ $\equiv$ \lum/L$_{\rm bol}$ = 0.043 $\times$ (L$_{\rm bol}$/10$^{45}$)$^{-0.357}$
(assuming  a  bolometric luminosity of
L$_{\rm bol}$($\approx$ L$_{\rm IR}$) $\approx$ 3 $\times$ 10$^{45}$ \ergs) for a typical quasar from \citet{pico07}, and the values of $r_{X,bol}$ calculated using the hard X-ray luminosity inferred for the CT and transmission scenario, respectively. 
In fact, the expected value of $r_{X,bol}$ = 0.03 is consistent with that derived assuming a \lum\ $>$  5 $\times$ 10$^{43}$ \ergs\ (estimated from the \lo\ luminosity and a CT absorber), i.e.  $r_{X,bol}$ $>$ 0.02, but it is much  higher than  $r_{X,bol}$ = 0.0004 derived for a \lum\ = 1.2 $\times$ 10$^{42}$ \ergs.

Detection of objects such as \irs\ is important because they provide useful templates to explore the multiwavelength properties of the  obscured accretion phenomenon without any luminosity bias.
The observed properties of \irs\ can be interpreted in the framework of  an evolutionary merger-driven scenario according to which a peculiar dust-cocooned, early stage in the life cycle of quasars is linked to a period of intense star-forming activity in the interacting host galaxy \citep{silk98,treister10}.
An easily observable outcome of this scenario  is indeed an enhancement of the IR luminosity, with the system undergoing a (U)LIRG phase powered both by the SB and the AGN. Furthermore, \citet{horellou95} measured a molecular hydrogen mass $M_{H2}$ of 4.1  $\times$ 10$^{9}$ $M_\odot$ for \irs\ that implies an  $L_{FIR}$/$M_{H2}$ ratio $\geq$ 100 $L_\odot$$M^{-1}_\odot$, i.e. a value typical for gas-rich mergers \citep{sanders91}.
According to model predictions, the SMBH at the center of \irs\ should be accreting close to the Eddington rate, in agreement with the quasar-like values of \lo(\lum) measured(estimated) for the AGN.

The interacting system IRAS 20210$+$1121, therefore,  surely deserves deeper investigations in the future in order to examine the possible presence of any structures associated with the merging process (i.e. outflows, inflows, obscured SB regions)  which could not be revealed by the observational data available so far, but  potentially  very useful for our understanding  of quasar evolution and AGN/SB triggering mechanisms.

\acknowledgements

EP, CV and SB acknowledge support under ASI/INAF contract I/088/06/0.
FN acknowledges support from the XMM-Newton-NASA grant NNX09AP39G.
Based on observations obtained with XMM-Newton, an ESA science mission with instruments and contributions directly funded by ESA Member States and NASA. 
%

{}
\end{document}